\PassOptionsToPackage{table}{xcolor}
\documentclass{article}
\usepackage{ijcai24}

\usepackage{times}
\usepackage{soul}
\usepackage{url}
\usepackage[hidelinks]{hyperref}
\usepackage[utf8]{inputenc}
\usepackage[small]{caption}
\usepackage{amsmath}
\usepackage{amsthm}
\usepackage{booktabs}
\usepackage{algorithm}
\newtheorem{theorem}{Theorem}

\usepackage{mdframed}

\newcommand{\citep}{\cite}
\newcommand{\citet}[1]{\citeauthor{#1} [\citeyear{#1}]}
\newcommand{\citepos}[1]{\citeauthor{#1}'s}
\newcommand{\citepossessive}[1]{\citeauthor{#1}'s (\citeyear{#1})}

\DeclareRobustCommand{\nUmErAL}[1]{}

\usepackage{booktabs}

\usepackage[inline]{enumitem}
\setlist{noitemsep}

\newcommand{\fbf}{\textbf}
\newcommand{\fsc}{\textsc}
\newcommand{\fsf}[1]{{\small\textsf{#1}}}
\newcommand{\frm}{\textrm}
\newcommand{\msf}{\mathsf}
\newcommand{\fsl}{\textsl}

\usepackage{amssymb}
\usepackage{mathrsfs}

\usepackage{mathpartir}
\newenvironment{inference}{%
\begin{mathpar}%
\mprset{sep=-2em,andskip=0pt,fraction={\cdot\cdots\cdot}}}{\end{mathpar}}

\newenvironment{communication}{\begin{mathpar}}{\end{mathpar}}
\newcommand{\aand}{\and\hspace*{-0.5em}}
\newcommand{\aaand}{\and\hspace*{-1em}}
\newcommand{\aaaand}{\and\hspace*{-1.4em}}

\usepackage{booktabs}
\usepackage{siunitx}
\usepackage{tikz}
\usepackage{standalone}
\usetikzlibrary{arrows}
\usetikzlibrary{calc}
\usetikzlibrary{arrows.meta}
\tikzstyle{syncaction}=[dashed,o-{Circle[]},>=stealth']

\usepackage{listings}

\usepackage{xcolor}
\definecolor{listbackgroundcolorlight}{rgb}{0.91,0.92,0.94}

\newcommand{\skey}{\ensuremath\msf{key}}
\newcommand{\swho}{\ensuremath{\msf{who}}}
\newcommand{\swhat}{\ensuremath{\msf{what}}}
\newcommand{\sdo}{\ensuremath{\msf{do}}}

\DeclareMathOperator{\nogo}{\msf{nogo}}
\DeclareMathOperator{\nono}{\msf{nono}}
\newcommand{\sayso}{\ensuremath{\msf{sayso}}}
\DeclareMathOperator{\attemptable}{\msf{attemptable}}
\DeclareMathOperator{\fits}{\msf{fits}}
\DeclareMathOperator{\abides}{\msf{abides}}
\DeclareMathOperator{\feasible}{\msf{feasible}}
\DeclareMathOperator{\jfeasible}{\msf{jointly-feasible}}
\DeclareMathOperator{\dominates}{\msf{dominates}}
\DeclareMathOperator{\unsocial}{\msf{unsocial}}
\DeclareMathOperator{\Soc}{\msf{S}}
\DeclareMathOperator{\Zed}{\msf{Z}}
\DeclareMathOperator{\Y}{\msf{Y}}

\DeclareMathOperator{\hampers}{\msf{hampers}}
\DeclareMathOperator{\denables}{\msf{enables}}

\newcommand{\lra}{\mbox{$\longrightarrow$}}  

\newcommand{\inn}{\ensuremath{\ulcorner\msf{in}\urcorner\,}}
\newcommand{\out}{\ensuremath{\ulcorner\msf{out}\urcorner\,}}
\newcommand{\nil}{\ensuremath{\ulcorner\msf{nil}\urcorner\,}}

\newcommand{\key}{\ensuremath{\ulcorner\msf{key}\urcorner}}

\newcommand{\mname}[1]{\fsl{#1}}
\newcommand{\pname}[1]{\fsl{#1}}
\newcommand{\rname}[1]{\textsc{#1}}
\newcommand{\paraname}[1]{\fsf{#1}}

\title{Langshaw: Declarative Interaction Protocols Based on Sayso and Conflict}
\author{%
Munindar P. Singh$^1$
\and
Samuel H. Christie V$^1$
\And
Amit K. Chopra$^2$
\affiliations
$^1$Department of Computer Science, North Carolina State University, Raleigh, NC 27695, USA\\
$^2$School of Computing and Communications, Lancaster University, Lancaster LA1 4WA, UK
\emails
mpsingh@ncsu.edu,
shcv@sdf.org,
amit.chopra@lancaster.ac.uk
}

\begin{document}
\maketitle

\begin{abstract}
Current languages for specifying multiagent protocols either over-constrain protocol enactments or complicate capturing their meanings. 
We propose Langshaw, a declarative protocol language based on (1) \emph{sayso}, a new construct that captures who has priority over setting each attribute, and (2) $\nono$ and $\nogo$, two constructs to capture conflicts between actions. 
Langshaw combines flexibility with an information model to express meaning. 
We give a formal semantics for Langshaw, procedures for determining the safety and liveness of a protocol, and a method to generate a message-oriented protocol (embedding needed coordination) suitable for flexible asynchronous enactment. 

\end{abstract}

\section{Introduction}

Our setting of interest is a \emph{decentralized} multiagent system (MAS) \citep{Boissier+23:Dagstuhl-report} in which agents, each reflecting its stakeholder's autonomy, interact in a loosely coupled manner. 
To achieve interoperation between agents independently of their construction or reasoning presupposes that we specify their interactions at a high level \citep{Poslad-07:FIPA}. 

Classically, agents interact by sending \emph{messages}, each a bundle of information, to one another. 
Achieving interoperation requires (see Section~\ref{sec:related-work}) specifying (1) the information each message conveys, (2) constraints on message ordering and occurrence, and (3) the meaning of each message in mental \citep{Sadek+97:ARTIMIS} or social \citep{Ijcai-99-ACL} constructs. 

However, modeling interaction via messages is problematic. 
First, a message is a low-level construct---the smallest operational unit of interaction. 
Models based on messages must contend with an explosion in possible messages and orders, and the complexity of coordination. 
Second, there is not always a natural mapping between messages and meanings. 

In contrast to existing MAS approaches, we model communicative actions directly inspired by \citepossessive{Austin62} speech act theory, separately both from messages and meanings. 

\paragraph{Approach.} 
We propose \emph{Langshaw} (after Austin's middle name), a declarative language for specifying protocols in terms of communicative actions. 
Langshaw enables the precise and succinct specification of protocols via novel primitives for expressing the information content and the social arrangements needed for coordination. 
Respecting the fundamental Austinian doctrine of \emph{saying makes it so}, Langshaw introduces \emph{sayso} to capture who can declare what information (i.e., ``make it so''). 
To support flexible interactions, the saysos of agents over the same information are prioritized. 
Langshaw also introduces primitives to directly capture \emph{conflict} between actions. 
Moreover, to support flexible interactions, Langshaw supports agents concurrently \emph{attempting} actions. 
The semantics ensures the consistency of the system state without unduly compromising flexibility.

\paragraph{Contributions.} 
One, we introduce the Langshaw language and give its semantics and efficient decision procedures for checking the safety and liveness of Langshaw specifications. 

Two, we bridge the gap between synchronous and asynchronous communication. 
Assuming synchronous actions simplifies specification and reasoning. 
We give Langshaw a synchronous semantics. 
However, synchrony is an unrealistic model for MAS intended to be deployed over asynchronous infrastructures such as the Internet. 
We show how to compile a Langshaw protocol into a protocol that can be enacted via asynchronous messaging. 
Our compilation highlights Langshaw's simplicity and high-level nature: A compiled protocol may involve subtle coordination that would be difficult to write by hand. 
We prove the correctness of the compiler. 

\section{Related Work and Novelty}
\label{sec:related-work}

Protocols are crucial to MAS engineering methodologies \citep{Cernuzzi+Zambonelli-04:Gaia,AOIS-05,Rooney+04:VIPER,Winikoff+Padgham-05}, but are traditionally expressed in semiformal notations such as AUML. 
Thus, they fall short of the goals for engineering MAS \citep{Winikoff-commitment-07}. 
Through its precision and flexibility, Langshaw offers an opportunity to rethink MAS methodologies by capturing stakeholder intuitions and helping realize implemented MAS. 

\citet{Ferrando+19:enactability} specify protocols as trace expressions over messages. 
They study diverse communication models, ranging from fully synchronous to fully asynchronous, but do not give a method to translate from one model to another, as we do. 
Moreover, as is common, \citeauthor{Ferrando+19:enactability} express the information content implicitly and coordination explicitly, which limits flexibility. 
\citet{Ferrando+17:DecAMon} devise techniques for the runtime verification of protocols; here, we focus on static verification.

BSPL (Blindingly Simple Protocol Language) \citep{AAMAS-BSPL-11,ICWS-LoST-11,AAMAS-BSPL-12} and HAPN (Hierarchical Agent Protocol Notation) \citep{Winikoff+18:HAPN} mix information content with coordination, gaining flexibility and concurrency at the cost of unwieldiness. 
HAPN's state-machine model is compatible with our synchronous semantics. 
BSPL supports flexible asynchronous enactments, and we show how to compile Langshaw into BSPL to benefit from asynchrony.

\citepossessive{Baldoni+14:JAAMAS} 2CL specifies social meaning via commitments and temporal and relational constraints. 
2CL does  not model information or support asynchrony. 
Langshaw supports encoding social meaning and thus helps support accountability \citep{Baldoni+23:accountability}. 
Its model enhances those of Custard \citep{AAMAS-16:Custard} and Clouseau \citep{AAAI-20:Clouseau} via primitives to capture coordination.

CArtAgO \citep{Ricci+09:CArtAgO} undergirds multiagent programming frameworks such as JaCaMo \citep{Boissier+13:JaCaMo}. 
CArtAgO agents coordinate via tuple spaces \citep{CarrieroGelernter89}. 
MAS programming models such as JaCaMo, Jason \citep{Vieira+07:speech-acts}, and Jade \citep{Bellifemine+07:JADE,Bergenti+20:JADE}
lack support for protocols. 
\citet{Ricci+19:scalable} argue for decentralization and scalability by combining MAS with the Web, but their interaction model is limited to HTTP (as on the Web). 
Langshaw give an abstract model for richer interactions, supporting asynchrony while based on a conceptually central social state (Section~\ref{sec:Langshaw-semantics-synchronous}).

Traditional approaches rely on a ``synchronizer'' in the infrastructure to decide which actions occurred in what order. 
For example, in \citepossessive{weyns:mas-model:2004} model, when agents perform conflicting actions simultaneously, the environment picks the successful ones. 
Such a synchronizer is not a social entity. 
It may avoid an integrity violation by forcing an ordering, but only arbitrarily. 
In Langshaw, such decisions are handled socially---a good protocol would have the right saysos to avoid the mishaps of concurrency.

\section{Langshaw Syntax and Informal Semantics}
\label{sec:Langshaw-semantics-synchronous}

We first conceive a Langshaw-based multiagent system as operating \emph{one} social artifact, the locus of its social state. 
Synchronous (and concurrent) social actions by (agents playing) roles in the system update this artifact. 
These updates respect local consistency (only different roles make take conflicting actions) and causality (an action may occur in a state only if the the information it relies on is present there). 
The social state is nothing but the set of such (performed) actions. 
The purpose of a Langshaw protocol is to specify such a system. 
Enacting the protocol means updating the social state. 

Listing~\ref{Langshaw:purchase} illustrates Langshaw's syntax (Table~\ref{syn:Langshaw}); we use it to explain how a protocol is enacted. 
As shown, the \mname{Purchase} protocol specifies roles \rname{seller}, \rname{buyer}, and \rname{shipper}. 

\begin{table}[htb]
\centering
\small
\begin{tabular}{l @{ }l@{ } l}\toprule

Protocol & $\lra$ & $\frm{Name}\ \frm{Roles}\ \frm{Attrs}\ \frm{Dos}\ \frm{Saysos}\ \lfloor \frm{Nogos} \rfloor\ \lfloor \frm{Nonos} \rfloor$ \\

Roles & $\lra$ & {\swho} $\mathscr{R}^+$\\

Attrs & $\lra$ & {\swhat} $\frm{Info}$\\ 

Info & $\lra$ & $[\mathscr{A}\ \skey]^+$ $[\frm{Expr}]^*$ \\
  
Expr & $\lra$ & $\mathscr{A}\ |\ \frm{Expr}\ \frm{and}\ \frm{Expr}\ |\ \frm{Expr}\ \frm{or}\ \frm{Expr}$\\

Dos & $\lra$ & {\sdo} $\frm{Action}^+$\\

Action & $\lra$ & $\mathscr{R}\ \colon\ [\mathscr{B}(\frm{Info})]^+$\\

Saysos & $\lra$ & {\sayso} $[\frm{Ranking}\ \colon\ \mathscr{A}]^+$\\

Ranking  & $\lra$ &  $\mathscr{R}\ |\   \mathscr{R}\ > \frm{Ranking}$ \\
  
Nogos & $\lra$ & $\nogo$ $[\frm{Action} \not\to \frm{Action}]^+$\\

Nonos & $\lra$ & $\nono$ $[\frm{Action}\ \frm{Action}]^+$\\

\bottomrule
\end{tabular}
\caption{Langshaw syntax. 
$\lra$, $|$, $*$, $+$ indicate production, choice, zero or more repetitions, and one or more repetitions, respectively. 
$[\,]$ and $\lfloor\,\rfloor$ indicate grouping and optionality, respectively. 
$\mathscr{A}$, $\mathscr{B}$, and $\mathscr{R}$ are, respectively, sets of (terminal) attributes, actions, and roles.}
\label{syn:Langshaw}
\end{table}

\begin{algorithm}[t]
\caption{Purchase in Langshaw.}
\label{Langshaw:purchase}
\begin{lstlisting}
Purchase |\label{line:name}| // Name of the protocol
 $\swho$ |\label{line:roles}| Buyer, Seller, Shipper // roles
 $\swhat$ |\label{line:completion}| ID $\skey$, Reject or Deliver // Completion
 $\sdo$ |\label{line:do}| // How to change the social state
  Buyer: RFQ(ID, item)
  Seller: Quote(ID, item, price)
  Buyer: Accept(ID, item, price, address)
  Buyer: Reject(ID, Quote)
  Seller: Instruct(ID, Accept, item, address, fee)
  Shipper: Deliver(ID, Instruct, item, address) |\label{line:Deliver}|
 $\sayso$ |\label{line:sayso}| // Social authority over what
  Buyer > Seller: item  |\label{line:item-sayso}|
  Seller > Buyer: price |\label{line:price-sayso}|
  Buyer: address
  Seller: fee
 $\nono$ |\label{line:nono}| // What pairs are incompatible
   Accept Reject |\label{line:conflict-accept-reject}|
   Reject Deliver |\label{line:conflict-deliver-reject}|
 $\nogo$ |\label{line:nogo}| // What actions prevent another
   Reject -/> Instruct |\label{line:nogo-reject-instruct}|
\end{lstlisting}
\end{algorithm}

Lines~\ref{line:do}--\ref{line:Deliver} specify social actions, each to be performed by a role; each action specifies one or more attributes (to capture its meaning), including one or more key attributes to uniquify instances. 
For example, \rname{buyer}  may attempt the action \mname{RFQ} by providing bindings (values) for \paraname{ID} and \paraname{item} and \rname{seller} may attempt \mname{Quote} by providing bindings for \paraname{ID}, \paraname{item}, and \paraname{price}. 
Each \mname{Action} is reified in an implicit \emph{action} attribute, which is bound if and only if that action has been instantiated for the specified key bindings. 
When an action includes such an attribute, it indicates a reliance on the first action. 
For example, \mname{Reject} applies to an instance of \mname{Quote}. 
\mname{Accept} could include \mname{Quote} but \paraname{item} and \paraname{price} make its meaning clearer.

Attributes are defined only in reference to their keys: it makes no sense to talk of an \paraname{item} without its \paraname{ID}. 
Therefore, we assume that if an attribute occurs in two actions, their keys overlap and their intersection uniquely determines that attribute. 
For a key attribute, a role may either generate a fresh binding or reuse a previous binding from the social state. 
For any other attribute, if it is bound in the social state relative to a key binding, a role must use the same binding relative to that key. 
 For example, in a social state with \mname{RFQ} with some \paraname{ID} and \paraname{item}, a \mname{Quote} or an \mname{Accept} with the same \paraname{ID} must contain the same \paraname{item}. 
If no such binding exists, the role may generate a fresh binding for that attribute only if it has \emph{sayso} over that attribute, as Line~\ref{line:sayso} and those following illustrate. 
For example, \rname{buyer} and \rname{seller} can both generate \paraname{item}. 
 
Attempts may be concurrent. 
When an agent attempts multiple actions, their bindings must be consistent with each other and the social state. 
For example, \rname{buyer} may concurrently attempt several \mname{RFQ}s, each with a distinct binding for \paraname{ID}. 
When multiple agents attempt actions concurrently, the agents need not be consistent with each other, just within each agent's actions and between the agent and the social state. 
For example, \rname{buyer} and \rname{seller} may concurrently attempt an \mname{RFQ} and \mname{Quote} by generating the same binding for \paraname{ID} but different bindings for \paraname{item} (since they both have sayso over \paraname{item}). 
In Langshaw, the saysos over the same attribute must be prioritized over agents (indicated by $>$): here, by Line~\ref{line:item-sayso}, \rname{buyer}'s sayso (being ranked higher) \emph{dominates}. 
Consequently, its attempt succeeds (updates the social state) whereas \rname{seller}'s attempt fails (is a noop). 

If only one agent attempts any actions, then all its attempts succeed and become part of the social state. 
If two or more agents attempt actions, exactly those of their collected attempts succeed where for each attribute not already bound in the social state, the attempting role has the highest sayso of all the roles concurrently attempting to produce a binding for that attribute. 
When attempts dominate each other (on different attributes), neither affects the social state. 
For example, if \rname{buyer} and \rname{seller} concurrently attempt \mname{Accept} and \mname{Quote} by each generating \paraname{item} and \paraname{price} for the same \paraname{ID}, both attempts fail because \rname{buyer}'s sayso dominates for \paraname{item} whereas \rname{seller}'s sayso dominates for \paraname{price}.

Line~\ref{line:nono} and those following specify pairwise conflicts between actions. 
\mname{Accept} and \mname{Reject} conflict, meaning that they are mutually exclusive for the same bindings of \paraname{ID}. 
But since they are both actions of \rname{buyer}, it can choose either one. 

Line~\ref{line:nogo} and those following specify an asymmetric conflict. 
Once \mname{Reject} is in the social state, \mname{Instruct} may not be performed. 
Importantly, \mname{Instruct} may precede or occur simultaneously with \mname{Reject}. 
This constraint is not needed in \pname{Purchase} but we insert it to explain the construct.

Consider \emph{safety}. Concurrent conflicting attempts by two agents can succeed, resulting in an inconsistent social state. 
A \emph{safe} protocol prevents such attempts. 
\mname{Purchase} is safe. 
\mname{Accept} and \mname{Reject} are both actions of \rname{buyer}, which means that if \rname{buyer} does one, it cannot do the other. 
\mname{Reject} and \mname{Deliver} are actions of \rname{buyer} and \rname{shipper}, respectively, but there is no social state in which both can be attempted concurrently.
To do \mname{Deliver}, \rname{shipper} needs to know the binding \paraname{address}, which can only happen if \rname{buyer} does \mname{Accept}, which rules out \mname{Reject} because it would be a local conflict at \rname{buyer}. 

Let protocol \mname{Unsafe Purchase} be obtained from Listing~\ref{Langshaw:purchase} by omitting Line~\ref{line:conflict-accept-reject}: thus, both \mname{Accept} and \mname{Reject} may occur. 
Suppose \mname{Accept} has occurred, followed by \mname{Instruct}. 
Now the social state is such that \mname{Reject} and \mname{Deliver} may both be attempted concurrently by \rname{buyer} and \rname{shipper}, respectively. 
Both \rname{buyer} and \rname{shipper} will succeed, thus violating the specified $\nono$ between \mname{Reject} and \mname{Deliver}. 

Consider \emph{liveness}. Line~\ref{line:completion} gives a \emph{completion} criterion for \mname{Purchase}'s enactments as a list of attributes or disjunctions of attributes; one or more of which are designated {\key} and distinguish enactments. 
Specifically, it says that an enactment of \mname{Purchase}, as identified by a binding for \paraname{ID}, is complete when a binding exists for either \mname{Reject} or \mname{Deliver}.

A protocol violates liveness if at least one of its enactments fails to complete. 
\mname{Purchase} has two alternative branches for an enactment, one ending with \mname{Reject} and the other ending with \mname{Deliver}. 
Thus, on each branch, \mname{Purchase}'s completion criterion is satisfied, which means \mname{Purchase} is live. 
Nonliveness could result from too little sayso or where the saysos induce a cyclic information dependency between the actions.



  

\section{Formal Semantics}
\label{sec:semantic-tableau-Langshaw}
Semantic tableaux \citep{Fitting-99:tableaux} are a computational representation for proofs. 
Each node is a \emph{social} state of the system, and the transitions are the successful concurrent actions attempted. 
Each tableau captures all possible enactments of a protocol: one per \emph{branch} beginning from the root. 
Our semantics (Figure~\ref{fig:semantics-synchronous}) is framed as inference rules to derive all possible transitions. 
Section~\ref{sec:reduced-tableaux} gives heuristics to produce a small tableau, i.e., a tractable model of a protocol.


\begin{figure}[t]
\centering
\begin{mdframed}
\begin{inference}
\inferrule*[LEFT=\fsc{Z-social }]%
{\Soc m_i[\vec{a_i}] \and m_i[\vec{a_i}] \fits m[\vec{a}]}
{\Zed(m[\vec{a}], \vec{a_i}\cap \vec{a})}
\end{inference}
\begin{inference}
\inferrule*[LEFT=\fsc{Z-subset }]%
{\Zed(m[\vec{a}], \vec{q_i}) \and \vec{q}\subseteq \vec{q_i}}
{\Zed(m[\vec{a}], \vec{q})}
\end{inference}
\begin{inference}
\inferrule*[LEFT=\fsc{Z-union }]%
{\{\Zed(m[\vec{a}], \vec{q_i})\}_{i=1}^n}
{\Zed(m[\vec{a}], \cup_{i=1}^n \vec{q_i})}
\end{inference}
%
%
\begin{inference}
\inferrule*[LEFT=\fsc{Attempt }]%
{\Zed(m[\vec{k},\vec{p}], \vec{q}) \aand \neg \Zed(m[\vec{k},\vec{p}], \vec{y}) \aand Y_x \vec{y}}
{\attemptable x\colon {m}[\vec{k},\vec{q}\cup\vec{y}]}
\end{inference}
\begin{inference}
\inferrule*[LEFT=\fsc{Abide }]%
{m_i[\vec{a_i}] \fits m[\vec{a}] \rightarrow m_i[\vec{p_i}\!\cap\! \vec{p}]\!=\!m[\vec{p_i}\!\cap\! \vec{p}]}
{m_i[\vec{a_i}] \abides m[\vec{a}]}
\end{inference}
\begin{inference}
\inferrule*[LEFT=\fsc{Unsocial-M }]%
{\Soc m_i[\vec{a_i}] \and \neg\, m_i \abides m}
{\unsocial m[\vec{a}]}
\end{inference}
\begin{inference}
\inferrule*[LEFT=\fsc{Unsocial-N }]%
{\Soc m_i[\vec{a_i}] \and m_i[\vec{a_i}] \fits m[\vec{a}] \\\\ \nogo(m_i, m) \lor \nono(m_i, m)}
{\unsocial m[\vec{a}]}
\end{inference}
\begin{inference}
\inferrule*[LEFT=\fsc{Feasible-1 }]%
{\attemptable m[\vec{a}] \aaand \neg \unsocial m[\vec{a}]}
{\feasible m[\vec{a}]}
\end{inference}
\begin{inference}
\inferrule*[LEFT=\fsc{Feasible }]%
{\{\feasible x_i\colon m_i[\vec{a_i}]\}_{i=1}^n\\\\
\wedge_{x_i=x_j} \neg\nono(m_i, m_j) \land m_i \abides m_j} 
{\feasible \{x_i\colon m_i[\vec{a_i}]\}_{i=1}^n}
\end{inference}
\begin{inference}
\inferrule*[LEFT=\fsc{Dominates }]%
{m_i[\vec{a_i}] \fits m[\vec{a}] \and p\in \vec{a_i}\cap \vec{a} \\\\ \neg\Zed (m[\vec{a}], p) \and Y_{x_i,x} p}
{x_i\colon\! m_i[\vec{a_i}] \dominates x\colon\! m[\vec{a}]}
\end{inference}
\begin{inference}
\inferrule*[LEFT=\fsc{Joint }]%
{\feasible \{m_i[\vec{a_i}]\}_{i=1}^n \aaaand \wedge_{i\neq j}\! \neg\, m_i \dominates m_j}
{\jfeasible \{x_i\colon m_i[\vec{a_i}]\}_{i=1}^n}
\end{inference}
\begin{communication}
\inferrule*[LEFT=\fsc{Communication }]%
{\jfeasible \{m_i[\vec{a_i}]\}_{i=1}^n}
{\Soc {x_i\colon m_i[a_i]}_{i=1}^n}
\end{communication}
\end{mdframed}
\caption{Synchronous semantics for Langshaw.}
\label{fig:semantics-synchronous}
\end{figure}

Figure~\ref{fig:semantics-synchronous} lists the inference rules for our semantics, based on propositional logic. 
Below, $m$ is an instance $x\colon m[\vec{a}]$ where $\vec{a}$ is its entire set of attributes. 
We write $x\colon m[\vec{k}, \vec{p}]$ or $m[\vec{k}, \vec{p}]$ to mean role $x$ performs $m$ with attributes $\vec{p}$ and key attributes $\vec{k}\subseteq \vec{p}$. 
We omit the role where it is clear. 
$\Soc$ captures the social state. 
$\Y_{x,y} a$ means $x$ has higher sayso over $a$ than $y$ (both $x$ and $y$ have sayso). 
$\Y_x a$ means $x$ has sayso over $a$ of whatever priority. 
Braces \{\,\} capture sets. 
Subscripts on sets and operators capture the obvious indices and ranges. 
An instance $m_i[\vec{k_i}, \vec{p_i}] \fits m[\vec{k}, \vec{p}]$ if and only if their common key attributes have the same bindings, i.e., $m_i[\vec{k_i} \cap \vec{k}]=m[\vec{k_i} \cap \vec{k}]$. 

In \fsc{Z-social}, $\Zed$ captures what bindings can be inferred from the social state relative to a (potential or actual) message instance. 
For an instance $m[\vec{k},\vec{p}]$, the bindings of its attributes occurring in a fitting instance in the present social state can be inferred (i.e., are known). 
\fsc{Z-subset} and \fsc{Z-union} state that $\Zed$ bindings are closed under subset and union. 

In \fsc{Attempt}, an instance is \emph{attemptable} by $x$ if (1) it includes any bindings already established (for the key) in the social state and (2) $x$ has sayso over the remaining attributes. 

In \fsc{Abide}, an instance \emph{abides} by another if: if the bindings of their common key attributes agree, then the bindings of all their common attributes also agree. 
Any $\nono$ and $\nogo$ constraints are ignored here and captured in \fsc{Feasible}. 

\fsc{Unsocial-M} states that an instance is \emph{unsocial} (i.e., socially inconsistent) if there is an instance in the social state with which it does  not abide. 
\fsc{Unsocial-N} states that an instance is \emph{unsocial} if an instance in the social state has a $\nono$ or a $\nogo$ (asymmetric) constraint toward that instance. 

\fsc{Feasible-1} states that an instance is \emph{feasible} if (1) it is attemptable and (2) its attribute bindings are not inconsistent with the social state. 
In \fsc{Feasible}, a set of instances is \emph{feasible} if each instance is feasible and for each pair of instances performed by the same role, they are not related by a nono and each instance in a pair abides by the other. 
This is a crucial design decision in Langshaw: Feasibility avoids only local incompatibilities. 
A protocol may be incorrect (e.g., unsafe) because of $\nono$ conflicts across roles. 
Since such errors cannot be avoided in a decentralized architecture, proper {\sayso}s are essential in a protocol that remains correct in asynchrony. 

\fsc{Dominates} captures that $m_i$ \emph{dominates} $m$ if they have the same bindings for their common key attributes, a common \emph{unbound} attribute $p$, and $x_i$ has {\sayso} priority on $p$ over $x$. 
(Thus, $m_i$ and $m$ can dominate each other.) 
Dominance goes away once the conflicting {\sayso} attribute ($y$ above) is bound. 

\fsc{Joint} states that a feasible set of instances whose members do not dominate another is \emph{jointly feasible}. 
If $m_i$ and $m$ dominate each other, at most one can be part of a jointly feasible set. 
Note that joint feasibility is closed under subsets. 

\subsection{Generating Enactments from a Tableau}

The preceding rules (dotted lines) show inference within a single social state: no $\Soc$ assertions are inferred, hence no state change. 
\fsc{Communication} (solid line) shows the progress of time. 
A set of jointly feasible actions may be concurrently performed and the resulting social state records each of the actions as having occurred via $\Soc$ assertions. 

A tableau begins from a root node---no $\Soc$ assertion. 
Any action there relies entirely on its performer's sayso. 
Under \fsc{Communication}, each jointly feasible set forks a tableau branch. 
Where a branch terminates (i.e., at a state with no more feasible actions) forms a unique history. 

\subsection{Verifying Correctness Properties}
\label{sec:verifying}
Properties of interest, e.g., liveness and safety, are concerned with the reachability or otherwise of a state through an enactment. 
These properties are expressed via propositional combinations of the actions (and attributes). 
We assert a property at the root. 
A consistent branch that ends provides an example of the property at the root. 
A branch that hits a contradiction is \emph{closed}; a tableau closes if all its branches close, which indicates the property is inconsistent and its negation is proved. 

For liveness, we derive a formula from its {\swhat} line---for the simple case, this means each protocol attribute becomes bound. 
To this end, we expand the social state to include bound attributes. 
For example, from $\Soc {\mname{Deliver}}$ we can infer $\Soc \mname{ID}$ and $\Soc \mname{item}$. 
A liveness violation occurs precisely when at least one of the attributes remains unbound. 
For example, in Listing~\ref{Langshaw:purchase}, $\neg \Soc {\mname{ID}} \lor (\neg \Soc {\mname{Reject}} \land \neg \Soc {\mname{Deliver}})$.
A consistent branch is a counterexample since our formula is negated. 
But if every branch closes, liveness is \emph{established}. 

For safety, an integrity violation is when two actions of different performers with a $\nono$ constraint occur. 
In Listing~\ref{Langshaw:purchase}, \mname{Reject} and \mname{Deliver} are such actions. 
A safety violation occurs for any such pair of actions. 
The negation of this property means that safety is preserved. 
We place the negated property, $\neg \Soc\text{\mname{Reject}} \lor \neg \Soc\text{\mname{Deliver}}$, at the root of the tableau. 
If any branch of the tableau closes, i.e., runs into a contradiction, we determine that safety is \emph{violated}. 

\begin{theorem}\label{thm:verification}
Liveness and safety properties based as above on the $\swhat$ and $\nono$ parts of a protocol are verified by negating and checking for contradiction. 
\end{theorem}
\noindent\emph{Argument}. 
Follows from the construction of the tableau. 

\section{Reducing the Tableau}
\label{sec:reduced-tableaux}
A tableau produced naively would include a branch for every jointly feasible set of actions at a state (node), leading to an exponential explosion of mostly redundant information. 
If we take the branches as individual actions, we will end up unrolling every feasible interleaving of the actions, also leading to an exponential explosion. 
How can we effectively reduce that redundancy? 
Our idea is to consolidate branches where possible and retain enough (heuristically, typically a few) branches to cover all semantically distinct possibilities.

\subsection{How Actions Relate in a Protocol}
Three cases are relevant. 
First, mutually \emph{unrelated} pairs of actions may occur in any combination or order. 
We can discard all but one branch in which they are performed: a single-shot concurrent performance would lead to a shallower tableau, but any arbitrary interleaving would produce the same social state. 
Second, where one action \emph{enables} another, the tableau would simply unfold such that an action would not become attemptable unless the right $\Soc$ assertions were present. 

Third, where two actions may interfere with each other, we must include enough branches in the tableau so that each alternative is included. 
We capture potential interference as \emph{hampers} below. 
If an action $m_1$ may disable $m_2$ (as based on a $\nono$ or a $\nogo$) or prevent its concurrent performance (as based on {\sayso} priority), we must include both possibilities with $m_1$ or $m_2$ going first and the other action being disabled or allowed later. 
We must recognize such cases even if one of the actions is not presently enabled and make sure we do not prematurely eliminate it. 
Figure~\ref{fig:enablement+interference} formalizes this reasoning. 

In \fsc{Enable}, action $x_i\colon m_i$ \emph{enables} action $x\colon m$ if $m_i$ is a potential precursor to $m$. 
That is, $x$ lacks sayso on some attribute $p$ of $m$ that is  not established from the social state (i.e., $\neg\Zed (m[\vec{a}], p)$) but $x_i$ has sayso on $p$ and $p$ appears in $m_i$. 
Action $m_i$ \emph{enables} action $m$ if there is a chain of one or more direct enablements from $m_i$ to $m$. 
\fsc{Chain} expresses that enablement is transitively closed. 

Action $m_i$ may \emph{hamper} action $m$ in two ways. 
The first case is if there is a $\nono$ or a $\nogo$ assertion involving $m_i$ and $m$. 
\mname{Accept} and \mname{Reject} and \mname{Reject} and \mname{Deliver} are examples. 
\fsc{Block} captures this case. 
The second case is when $x_i$ has priority over $x$ for an attribute $p$ that is not yet established in the social state, exactly as specified by $\dominates$. 
\mname{Quote} and \mname{Accept} is an example pair. 
\fsc{Delay} captures this case. 

\fsc{Future} extends \emph{hampers} to accommodate future (potential) interference. 
Action $m_i$ \emph{hampers} action $m$ if $m_i$ hampers action $m_j$ and $m$ enables $m_j$.

\begin{figure}[t]
\centering
\begin{mdframed}
\begin{inference}
\inferrule*[LEFT=\fsc{Enable }]%
{m_i[\vec{a_i}] \fits m[\vec{a}] \and \neg\Zed (m[\vec{a}], p)\\\\
\Y_{x_i} p \and \neg\! \Y_x p \and p\in \vec{p_i}\cap\vec{p}}
{x_i\colon m_i[\vec{a_i}] \ \denables\  x\colon m[\vec{a}]}
\end{inference}
\begin{inference}
\inferrule*[LEFT=\fsc{Chain }]%
{m_i[\vec{a_i}] \ \denables\  m[\vec{a}] \aand m[\vec{a}] \ \denables\  m_j[\vec{a_j}]}
{m_i[\vec{a_i}] \ \denables\  m_j[\vec{a_j}]}
\end{inference}
\begin{inference}
\inferrule*[LEFT=\fsc{Block }]%
{m_i[\vec{a_i}] \fits m[\vec{a}] \\\\ \nono(m_i, m) \lor \nogo(m_i, m)}
{x_i\colon m_i[\vec{a_i}] \ \hampers\ x\colon m[\vec{a}]}
\end{inference}
\begin{inference}
\inferrule*[LEFT=\fsc{Delay }]%
{x_i\colon m_i[\vec{a_i}] \ \dominates\ x\colon m[\vec{a}]}
{x_i\colon m_i[\vec{a_i}] \ \hampers\ x\colon m[\vec{a}]}
\end{inference}
\begin{inference}
\inferrule*[LEFT=\fsc{Future }]%
{m_i[\vec{a_i}] \hampers m_j[\vec{a_j}] \aaand m[\vec{a}] \denables m_j[\vec{a_j}]}
{m_i[\vec{a_i}] \ \hampers\  m[\vec{a}]}
\end{inference}
\end{mdframed}
\caption{Reasoning about enablement and interference.}
\label{fig:enablement+interference}
\end{figure}

\subsection{Computing Jointly Feasible Sets of Actions}

Identify all feasible actions at a state. 
Next, group these actions into jointly feasible sets, reflecting two intuitions. 

First, [Colors] when an action hampers another, the tableau must include both orders and maximally group nonhampering actions to reduce its size. 
To this end, create a graph whose vertices are the feasible actions and an edge indicates one $\hampers$ another. 
A \emph{coloring} \citep{Brelaz-79:graph-coloring} finds sets of vertices with no edge between them. 
Each set is jointly feasible (no action in a set $\hampers$ another), but maximality maps to minimum graph coloring, which is NP-Hard. 
We apply \citepos{Brelaz-79:graph-coloring} polynomial-time approximation to compute Colors. 

Second, [Concs] when an action hampers another, the two can be performed concurrently if they have different performers, there is a $\nono$ between them ($\nogo$s are OK), and neither performer dominates the other on some unbound attribute that is common to the actions. 
This gives the set Concs.



We construct a \emph{reduced tableau} by including branches whose action sets are in Colors $\cup$ Concs.

\begin{theorem}\label{thm:reduced}
A reduced tableau is closed for a property constructed from Boolean combinations of $\Soc$ assertions if and only if the full tableau is closed for that property. 
\end{theorem}
\noindent\emph{Argument}. 
\fbf{If:} All branches in a reduced tableau are jointly feasible and occur in the full tableau. 
\fbf{Only if:} A branch closes based on the $\Soc$ assertions on it. 
Since the $\nogo$ and $\sayso$ constraints are incorporated into the tableau construction of Figure~\ref{fig:semantics-synchronous}, the only closures (contradictions) observed on a branch pertain to $\nono$ constraints. 
Suppose the reduced tableau lacks an equivalent branch. 
By Concs, it must include branches for the $\nono$ constraints, let $\Soc m$ be the first assertion that it does not include. 
Therefore, it must include an action $\Soc m'$ that disables $\Soc m$. 
By our coloring construction, that means $\Soc m$ should have been in a different color from $\Soc m'$. 
Therefore, there is a branch including $\Soc m$ on which no $\Soc m'$ precedes it. 
Hence, we have a contradiction. 

Theorem~\ref{thm:reduced} is important because it means all verification of properties based on $\Soc$ (including safety and liveness, as seen in Section~\ref{sec:verifying}) can be carried out on a reduced tableau.

\section{Compilation into Asynchronous Protocols}
\label{sec:BSPL-semantics-asynchronous}
For concreteness, we adopt BSPL for specifying asynchronous messaging protocols. 
As Listing~\ref{bspl:Accept-Reject} shows, a BSPL protocol lists roles, a completion criterion, and one or more messages. 
Each message specifies information dependencies via adornments \inn, \out, and \nil. 
Each role has a local state and may send a message only if it is compatible with its local state. 
Specifically, let $m$ be a message schema $x\mapsto y\colon m[\vec{p}_I, \vec{p}_O, \vec{p}_N]$, where $\vec{p}_I, \vec{p}_O, \vec{p}_N$ are sets of its parameters adorned \inn, \out, and \nil, respectively. 
An instance of $m$ is a message that has bindings only for the \inn and \out parameters; $x$ may send an instance of $m$ only if the bindings of the \inn parameters are known (already present in its local state), bindings of the \out parameters are unknown (but added to the local state upon sending), bindings of the \nil parameters are unknown (and not added to the local state). 
For example, to send an \mname{Accept}, \rname{buyer} must know \paraname{ID} and \paraname{item} and not know \paraname{price}. 
Moreover, \mname{Accept} and \mname{Reject} are mutually exclusive because each includes \paraname{done} as \out. 

\begin{algorithm}[t]
\caption{A simple protocol to explain BSPL constructs.}
\label{bspl:Accept-Reject}
\begin{lstlisting}
Accept-Reject //Protocol name
 roles B, S //roles Buyer and Seller
 parameters $\msf{out}$ ID $\msf{key}$, $\msf{out}$ done //completion criterion
 S $\mapsto$ B: Offer[$\msf{out}$ ID $\msf{key}$, $\msf{out}$ item, $\msf{out}$ price] // Message
 B $\mapsto$ S: Accept[$\msf{in}$ ID $\msf{key}$, $\msf{in}$ item, $\msf{in}$ price, $\msf{out}$ done]
 B $\mapsto$ S: Reject[$\msf{in}$ ID $\msf{key}$, $\msf{in}$ item, $\msf{in}$ price, $\msf{out}$ done]
\end{lstlisting}
\end{algorithm}

Figure~\ref{fig:BSPL-asynchronous} shows \citepossessive{IJCAI-21:Tango} tableau-based asynchronous semantics of BSPL. 
Each tableau branch (sequence of message send and receive events) is an enactment. 
$K_x$ captures that $x$ knows specified parameter bindings. 
$L_x$ means $x$ sent or received a message, simulating a channel. 
Initially, each role knows no bindings. 
What is known to each role grows monotonically: the bindings are immutable, and each message sent and received adds to a role's knowledge. 

\begin{figure}[t]
\centering
\begin{mdframed}
\begin{mathpar}
\inferrule*[LEFT=\fsc{Send }]%
  {K_x \vec{p}_I \and \neg K_x \vec{p}_O \and \neg K_x \vec{p}_N}
  {L_x (x\mapsto y\colon m[\vec{p}_I, \vec{p}_O, \vec{p}_N]) \and K_x \vec{p}_O}
\end{mathpar}
\begin{mathpar}
  \inferrule*[LEFT=\fsc{Receive }]%
  {L_x (r = x\mapsto y\colon m[\vec{p}_I, \vec{p}_O, \vec{p}_N])}
  {L_y r \and K_y \vec{p}_I \and K_y \vec{p}_O}
\end{mathpar}
\end{mdframed}
\caption{Asynchronous semantics for BSPL.}
\label{fig:BSPL-asynchronous}
\end{figure}

In \fsc{Send}, an instance of $x\mapsto y\colon m[\vec{p}_I, \vec{p}_O, \vec{p}_N]$ is enabled for emission by $x$ if and only if $x$ knows $\vec{p}_I$ but does not know $\vec{p}_O$ or $\vec{p}_N$. 
Concomitantly with the emission, the sender produces and comes to know the bindings for $\vec{p}_O$. 
%
In \fsc{Receive}, a message from $x$ to $y$ is enabled for reception by $y$ if and only if $x$ has (previously) sent that message. 
Concomitantly with reception, $y$ comes to know the bindings for $\vec{p}_I$ and $\vec{p}_O$.

Each tableau branch takes up one of the allowed observations (emission or reception of a message) and maps to an enactment. 
A tableau contains all possible observation orders. 

\subsection{Compiling Langshaw}
\label{sec:Langshaw-compilation}

We now describe how to compile Langshaw specifications into BSPL in a manner that respects the demands of asynchrony. 
Langshaw roles map to BSPL roles, attributes to parameters, actions to messages, saysos to data flows, and nonos to integrity constraints. 
Resolving the saysos and nonos requires parameters and messages not present in Langshaw. 

\paragraph{Completion Requirements.} Langshaw completion requirements are a sequence of disjunctive clauses. 
Each attribute (atom in a clause) yields a BSPL parameter of the same name. 
Each disjunction is replaced by a parameter representing the achievement of any of its terms. 
When generating protocol messages, we append messages that are enabled when any of the clause terms is completed to notify each role that the clause is satisfied. 

Listing~\ref{lst:purchase-completion} (full version online; URL below) shows some lines generated from \pname{Purchase}. 
The $\swhat$ line yields two clauses. 
ID maps to BSPL parameter \paraname{ID}. 
And, Reject or Deliver maps to \paraname{done0}. 
Each role gets a message for each of its actions. 
We also create messages to convey important facts to other roles: e.g., \mname{Reject} maps to messages indicating that the primary \mname{Reject} has been sent and the $\swhat$ line satisfied. 

\begin{algorithm}[t]
\caption{Purchase: Completion requirements.}
\label{lst:purchase-completion}
\begin{lstlisting}
 parameters $\msf{out}$ ID $\msf{key}$, $\msf{out}$ done0
 ... 
 Buyer $\mapsto$ Seller: Reject#done0[$\msf{in}$ ID $\msf{key}$, $\msf{in}$ Reject, $\msf{out}$ done0]
\end{lstlisting}
\end{algorithm}

\paragraph{Generating Messages from Actions.}
\label{sec:action-messages}
Each action yields a BSPL message schema, whose sender is the performer and whose receiver is a role that needs the content of the message. 
\begin{itemize}
\item Each attribute (including the attributes reifying the actions) becomes a message parameter. 
\item The adornment of each parameter depends on whether the action's role (message's sender) has {\sayso}. 
\item Each combination of parameter adornments yields a different \emph{morph}, i.e., message schema variant. 
\item The above possibilities are constrained via a series of filters to remove incorrect and redundant combinations. 
\end{itemize}

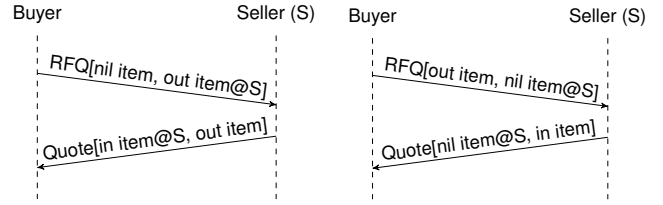
\begin{figure}[t]
\centering
\resizebox{0.48\columnwidth}{!}{\begin{tikzpicture}%
\usetikzlibrary{arrows}
\usetikzlibrary{calc}
\tikzstyle{role}=[thin,draw,align=center,font=\sffamily,rectangle,anchor=center,minimum height=5ex,minimum width=6ex,inner sep=1]

\tikzstyle{m_label_base}=[draw=none,midway,fill=none,sloped,align=center,font=\sffamily]
\tikzstyle{m_label_up}=[m_label_base,above=-2pt]

\tikzstyle{message}=[->, >=stealth']

\tikzstyle{emptybox}=[draw=none,minimum height=3ex]

\matrix[ampersand replacement=\&] () [row sep=10,column sep=120] {
  \node[emptybox] (a) {};
  \&  \node[emptybox] (b) {};\\
  \node (a-zero) {};
  \& \node (b-zero) {};\\
  \node (a-one) {};
  \& \node (b-one) {};\\
  \node (a-two) {};
  \& \node (b-two) {};\\
  \node (a-three) {};
  \& \node (b-three) {};\\
  \node (a-last) {};
  \& \node (b-last) {};\\
};

\node [role,draw=none,anchor=south] at (a) {Buyer};
\node [role,draw=none,anchor=south] at (b) {Seller (S)};

\draw [dashed] (a.center)--(a-last.center);
\draw [dashed] (b.center)--(b-last.center);

\draw [message] (a-zero.center)--node [m_label_up] {RFQ[nil item, out item@S]} (b-one.center);

\draw [message] (b-two.center)--node [m_label_up] {Quote[in item@S, out item]} (a-three.center);

\end{tikzpicture}
}~~
\resizebox{0.48\columnwidth}{!}{
\begin{tikzpicture}%
\usetikzlibrary{arrows}%
 \usetikzlibrary{calc}
  \tikzstyle{role}=[thin,draw,align=center,font=\sffamily,rectangle,anchor=center,minimum height=5ex,minimum width=6ex,inner sep=1]

  \tikzstyle{m_label_base}=[draw=none,midway,fill=none,sloped,align=center,font=\sffamily]
  \tikzstyle{m_label_up}=[m_label_base,above=-2pt]

\tikzstyle{message}=[->, >=stealth']

\tikzstyle{emptybox}=[draw=none,minimum height=3ex]

\matrix[ampersand replacement=\&] () [row sep=10,column sep=120] {
  \node[emptybox] (a) {};
  \&  \node[emptybox] (b) {};\\
  \node (a-zero) {};
  \& \node (b-zero) {};\\
  \node (a-one) {};
  \& \node (b-one) {};\\
  \node (a-two) {};
  \& \node (b-two) {};\\
  \node (a-three) {};
  \& \node (b-three) {};\\
  \node (a-last) {};
  \& \node (b-last) {};\\
};

\node [role,draw=none,anchor=south] at (a) {Buyer};
\node [role,draw=none,anchor=south] at (b) {Seller (S)};

\draw [dashed] (a.center)--(a-last.center);
\draw [dashed] (b.center)--(b-last.center);

\draw [message] (a-zero.center)--node [m_label_up] {RFQ[out item, nil item@S]} (b-one.center);

\draw [message] (b-two.center)--node [m_label_up] {Quote[nil item@S, in item]} (a-three.center);

\end{tikzpicture}
}
\caption{Sayso delegation pattern showing alternative enactments.}
\label{fig:sayso-pattern-delegate}
\end{figure}

\paragraph{Modeling Sayso in BSPL.} We model {\sayso} in BSPL via \emph{delegation}. 
That is, since \rname{Buyer} has priority over \rname{Seller} to bind \paraname{item}, it must either bind \paraname{item}, or bind a newly added \emph{delegation parameter} (\paraname{item@Seller}) to empower \rname{Seller} to bind it. 
Conversely, \rname{Seller} cannot bind \paraname{item} until \rname{Buyer} has delegated that authority. 
Figure~\ref{fig:sayso-pattern-delegate} illustrates the delegation pattern with message sequence diagrams. 
The pattern is inherently asymmetric: traditionally the priorities are arbitrary \citep{Mattern-90:termination} whereas our priorities have a social basis.

\paragraph{Filtering Out Improper Schemas.}
The above steps produce redundant or incorrect morphs. 
We apply a series of filters to exclude such cases. 
For instance, reasoning about causality is needed to ensure enactability and reasoning about priority to avoid race conditions. 

\paragraph{Conflicts and Data Flow.}
To generate BSPL messages, we model a $\nono$ constraint by adding \nil parameters to each of the conflicting messages. 
For example, in \pname{Purchase}, the \mname{Reject} and \mname{Deliver} actions conflict, so we add $\msf{nil}$ \texttt{Deliver} to each generated \mname{Reject} morph, and $\msf{nil}$ \texttt{Reject} to each generated \mname{Deliver} morph, as shown in Listing~\ref{lst:reject-deliver-conflict}. 
Likewise, we model a $\nogo$ constraint ($a\not\to b$) by placing a \nil parameter on the message for the disabled action, i.e., $b[\ldots \nil a]$. 
This approach echoes the idea of disabling an action until information is received to proceed \citep{Icde96}.

\begin{algorithm}[t]
\caption{Realizing the Reject and Deliver conflict.}
\label{lst:reject-deliver-conflict}
\begin{lstlisting}
 B $\mapsto$ S: Reject[$\msf{in}$ ID $\msf{key}$, $\msf{in}$ Quote, $\msf{out}$ Reject, $\msf{nil}$ Accept, |\textbf{$\msf{nil}$~Deliver}|]
 Sh $\mapsto$ S: Deliver[$\msf{in}$ ID $\msf{key}$, $\msf{in}$ Instruct, $\msf{in}$ item, $\msf{in}$ address, $\msf{out}$ Deliver, |\textbf{$\msf{nil}$~Reject}|]
\end{lstlisting}
\end{algorithm}

Next, we model the protocol's data flow by specifying the recipients of each message schema. 
A role can observe any attribute (including an action) that features in an action it can perform and any attribute present in the $\swhat$ line of the protocol. 
Information is shared at the granularity of actions, not attributes piecemeal, so we determine which roles can see what action. 
We generate a separate message schema for each of the desired data flows. 
Since BSPL does not support multicast, we emulate any multicast with separate messages. 
The first message schema (to one role) has the relevant parameters as \out; message schemas to other roles include the same parameters as \inn. 
As described above, every message is given a least one \out parameter. 

\begin{theorem}\label{thm:compilation}
For any asynchronous messaging enactment produced by a BSPL protocol generated through our method, there is a corresponding synchronous enactment produced by the Langshaw semantics. 
\end{theorem}

\noindent\emph{Argument}. 
We derive messages using Langshaw's semantics. 
Each action maps to a message and each attribute to a parameter with the constraints ($\nono$, $\nogo$, and $\sayso$ dominance) via crafted \nil parameters. 
We filter the schemas according to rules derived from the delegation model of {\sayso}, i.e., constraining the available actions according to \fsc{Dominates}. 
Thus, the generated BSPL protocol allows moves corresponding to any Langshaw action. 
Due to our filters, it cannot make any moves prevented by the {\sayso} constraints. 
Although Langshaw allows concurrent actions and BSPL does not allow multicast, the effect of multicast is achieved through additional messages: such messages can be delayed, as allowed by the BSPL's asynchronous semantics, but not disabled. 
We capture Langshaw's disjunctive completion requirements via messages producing a designated parameter. 

\subsection{Empirical Results}

We implemented our verifier for Langshaw protocols in Python. 
Table~\ref{tab:results} shows the performance (averaged over 10 runs) of testing liveness (including the time for constructing a tableau) for several protocols from the literature. 
The times for safety are similar, though slightly faster in most cases.

\newcommand{\colHleft}[1]{\multicolumn{1}{l}{\textbf{#1}}}
\newcommand{\colH}[1]{\multicolumn{1}{r}{\textbf{#1}}}
\begin{table}[htb]
\centering
\begin{tabular}{l S S S}\toprule
\colHleft{Protocol}         & \colH{Nodes} & \colH{Branches} & \colH{Time (ms)} \\ \midrule
\mname{Redelegation}       & 4     & 1        & 3.5   \\
\mname{Unsafe Purchase}    & 49.5  & 19.9     & 901.5 \\
\mname{PO-\ldots-Ship} & 12.3  & 2.4      & 92.6  \\
\mname{Either-Offer}       & 3     & 1        & 1.5   \\
\mname{Refund}             & 12.2  & 4.6      & 99.0  \\
\mname{Purchase}           & 28.2  & 8.1      & 480.1 \\
\mname{Unsafe}             & 5     & 3        & 3.2   \\
\mname{Block-Contra}       & 3     & 1        & 0.8   \\
\mname{Nonlive}            & 1     & 1        & 0.2   \\
\mname{CompositeKey}       & 3     & 1        & 2.5   \\
\mname{RFQ-Quote}          & 6     & 2        & 6.5   \\
\mname{Rescind}            & 7.5   & 2.5      & 23.7  \\
\mname{Block-Contra v2}    & 17    & 8        & 25.8  \\

\bottomrule
\end{tabular}
\caption{Statistics of Langshaw protocol liveness verification. \mname{Purchase}, \mname{Unsafe Purchase}, and \mname{Nonlive} are as specified above. The remaining protocols are in the online supplement (URL below). 
Time to verify is given in milliseconds. 
All except \mname{Unsafe Purchase} and \mname{Unsafe} were safe; all except \mname{Nonlive} were live. 
Our experimental rig was an ASUS Zenbook S13 with an AMD Ryzen 7 6800U CPU and 16GB of LPDDR5 RAM, running Linux.
}
\label{tab:results}
\end{table}

The results show that the Langshaw verifier is effective. 
The node and branch numbers are sometimes fractions, due to unordered sets in the implementation randomizing the selection and thus producing different numbers of nodes and branches in different runs of the verifier. 

\section{Discussion: Conclusion and Perspectives}

Specifying coordination constraints for MAS can be nontrivial. 
Langshaw's abstractions force a designer to think of coordination in social terms. 
The synchronous semantics provides a simplified model for a MAS engineer while maintaining the realism and power of asynchrony in that conflicting action attempts by multiple agents may succeed though they violate a $\nono$ constraint. 
Thus, it does  not obviate correctness concerns arising in asynchrony and thereby facilitates translation to an asynchronous protocol. 

For now, the best approach to implement agents to participate in Langshaw protocols is to (1) verify a Langshaw protocol; (2) compile it to BSPL; and (3) apply a BSPL programming model such as Kiko \citep{AAMAS-23:Kiko}. 
Research into native programming models for Langshaw is needed. 
Both BDI-based and the newer hypermedia-based programming models \citep{Vachtsevanou+23:signifiers} are promising as direct programming models for Langshaw because they naturally complement Langshaw's information orientation. 

The engineering MAS community has long debated the relative merits of synchrony and asynchrony \citep{IC-col-3:2}. 
Langshaw bypasses some of the debate by showing that a synchronous semantics can be a pathway to asynchrony. 
Popular approaches provide constructs such as artifacts \citep{Ricci+09:CArtAgO} or the environment \citep{Weyns+07:environment} that provide a unitary view of the state of a MAS. 
Though they may be accessed through low-level asynchronous means, their unitary view reflects a central point. 
Blockchain provides a shared state between otherwise independent entities \citep{Mendling+18:Blockchain-BPM}. 
Langshaw's synchronous semantics provides a basis for interaction governed by sayso, which is more conducive to business meaning \citep{IC-21:Hercule} than arbitrary ordering, and enables maximal concurrency given a protocol.

Theoretical studies of protocols, e.g., \citep{JAAMAS-Algebra-07,Ferrando+19:enactability}, may need to be revisited in light of Langshaw. 
Further, the problem of verifying protocol-based agents \citep{baldoni:conformance:2006:icsoc} remains largely unaddressed for information-based representations.

\paragraph{Online supplement.}
The code and all examples are available online at
\url{https://gitlab.com/masr/langshaw}

\section*{Acknowledgments}
Thanks to the anonymous reviewers for helpful comments. Thanks to the NSF (grant IIS-1908374) and EPSRC (grant EP/N027965/1) for support.

\bibliographystyle{named}
\DeclareRobustCommand{\nUmErAL}[1]{#1}\DeclareRobustCommand{\nAmE}[3]{#3}\DeclareRobustCommand{\nUmErAL}[1]{#1}\DeclareRobustCommand{\nAmE}[3]{#3}

\end{document}